\documentclass{tlp}

\usepackage{amsmath,amssymb,graphicx}

\renewcommand{\geq}	{\geqslant}
\renewcommand{\emptyset}{\varnothing}

\newtheorem{theorem}{Theorem}[section]
\newtheorem{definition}[theorem]{Definition}
\newtheorem{example}[theorem]{Example}
\newtheorem{lemma}[theorem]{Lemma}
\newtheorem{note}[theorem]{Note}

\newcommand{\prop}	{\textit{prop}}
\newcommand{\friends}	{\ensuremath{\mathit{friends}}}
\newcommand{\obviated}	{\ensuremath{\mathit{obviated}}}
\newcommand{\update}	{\ensuremath{\mathit{update}}}
\newcommand{\holds}[2]	{\ensuremath{\mathit{holds}({#1},{#2})}}
\newcommand{\eclipse}	{ECL$^i$PS$^e$}

\newcommand{\A}		{\ensuremath{\ \wedge\ }}
\newcommand{\Or}	{\ensuremath{\ \vee\ }}
\newcommand{\fa}	{\ensuremath{\forall}}
\newcommand{\te}	{\ensuremath{\exists}}

\newcommand{\LL}	{\ensuremath{\ldots}}
\newcommand{\ES}	{\ensuremath{\emptyset}}
\newcommand{\sse}	{\ensuremath{\subseteq}}
\newcommand{\po}	{\ensuremath{\sqsubseteq}}
\newcommand{\ra}	{\ensuremath{\rightarrow}}
\newcommand{\C}[1]	{\ensuremath{\{{#1}\}}}
\newcommand{\p}[2]	{\ensuremath{\langle #1 \ ; \ #2 \rangle}}
\newcommand{\NI}	{\noindent}
\newcommand{\HB}	{\hfill{$\Box$}}

\newcommand{\III}	{\vspace{3 mm}}

\def\smallromani{\renewcommand{\theenumi}{\roman{enumi}}
\renewcommand{\labelenumi}{(\theenumi)}}

\newcommand{\Proof}{\NI{\bf Proof.}\ }

\title{Schedulers and Redundancy for a Class of Constraint Propagation Rules}

\author{Sebastian Brand and Krzysztof R. Apt}

\submitted{31 August 2002}
\revised{7 November 2003, 23 April 2004}
\accepted{9 August 2004}

\begin{document}
\maketitle

\begin{abstract}
  We study here schedulers for a class of rules that naturally arise
  in the context of rule-based constraint programming. We
  systematically derive a scheduler for them from a generic iteration
  algorithm of \cite{Apt00a}.  We apply this study to so-called
  membership rules of \cite{AM01}.  This leads to an
  implementation that yields a considerably better performance
  for these rules than their execution as standard \texttt{CHR} rules.
  Finally, we show how redundant rules can be identified and how
  appropriately reduced sets of rules can be computed.
\end{abstract}


\section{Introduction}
\label{sec:introduction}

In this paper we identify a class of rules that naturally arise in the
context of constraint programming represented by means of rule-based
programming and study efficient schedulers for these rules.  We call
these rules \emph{propagation} rules, in short {\prop} rules.  An important class of {\prop} rules are
the \emph{membership rules}, introduced in \cite{AM01}.
An example of a membership rule is
\[
        x \in \{3,4,8\}, y \in \{1,2\} \ra z \neq 2.
\]
Informally, it should be read as follows: if the domain of $x$ is included in
\{3,4,8\} and the domain of $y$ is included in \{1,2\}, then 2 is removed from the
domain of $z$.

In the computations of constraint programs the variable domains
gradually shrink.  So if the domain of $x$ is included in
\{3,4,8\}, then it will remain so during the computation. In turn, if
2 is removed from the domain of $z$, then this removal operation does
not need to be repeated.  The {\prop} rules generalize these
observations to specific conditions on the rule condition and body.

In the resulting approach to constraint programming the computation
process is limited to a repeated application of the {\prop} rules
intertwined with splitting (labeling).  So the viability of this
approach crucially depends on the availability of efficient schedulers
for such rules.  This motivates the work here reported.  We provide an
abstract framework for such schedulers and use it as a basis for an
implementation.

More precisely, to obtain appropriate schedulers for the {\prop} rules
we use the generic approach to constraint propagation algorithms
introduced in \cite{Apt99b} and \cite{Apt00a}.  In this
framework one proceeds in two steps.  First, a generic iteration
algorithm on partial orderings is introduced and proved correct in an
abstract setting. Then it is instantiated with specific partial
orderings and functions to obtain specific constraint propagation
algorithms.  In this paper, as in \cite{Apt00a}, we take into
account information about the scheduled functions, which are here the
{\prop} rules. This yields a specific scheduler in the form
of an algorithm called \texttt{R}.

We then show by means of an implementation how this abstract framework
can be used to obtain a scheduler for the membership rules.  The
relevance of the membership rules for constraint satisfaction problems
(CSPs) with finite domains stems from the following observations made
in \cite{AM01}:
\begin{itemize}

\item constraint propagation can be naturally achieved by repeated
  application of the membership rules;
  
\item in particular the notions of arc consistency and hyper-arc
  consistency can be characterized in terms of the membership rules;

\item for constraints explicitly defined on small finite domains all
  valid membership rules can be automatically generated; (For a
  more referent work on the subject of an automatic generation of such rules
  see \cite{AR01}.)

\item many rules of the \texttt{CHR} language (Constraint Handling
  Rules) of \cite{FruehwirthJLP98} that are used
  in specific constraint solvers are in fact membership rules.
  In the logic programming approach to constraint programming
  \texttt{CHR} is the language of choice to write constraint solvers.
\end{itemize}

The implementation is provided as an \eclipse{} program that accepts a set
of membership rules as input and constructs an \eclipse{} program that
is the instantiation of the \texttt{R} algorithm for this set of rules.  Since
membership rules can be naturally represented as \texttt{CHR} propagation
rules, one can assess this implementation by comparing it with the
performance of the standard implementation of membership rules in the
\texttt{CHR} language. By means of various benchmarks we found that our
implementation is considerably faster than \texttt{CHR}.
It is important to stress that this implementation was obtained by
starting from ``first principles'' in the form of a generic iteration
algorithm on an arbitrary partial ordering.  This shows the practical
benefits of studying the constraint propagation process on an abstract
level.

Additionally, we clarify how to identify {\prop} rules that are
redundant for the considered computations and how to compute
appropriately reduced sets of rules.  The concept of redundancy is
formalized here in a ``semantic'' sense that takes into account the
type of computations performed by means of the considered rules. We
provide a simple test for redundancy that leads to a natural way of
computing minimal sets of rules in an appropriate sense.  The
computation of a specific minimal set for the membership rules is then
implemented in \eclipse.

\texttt{CHR} is available in a number of languages including the
\eclipse{} and the Sicstus Prolog systems. In both cases \texttt{CHR}
programs are compiled into the host language, so either \eclipse{}
or the Sicstus Prolog.  There is also a recent implementation in Java,
see \cite{AKSS01}.  To make \texttt{CHR} usable it is important that
its implementation does not incur too much overhead.  And indeed a
great deal of effort was spent on implementing \texttt{CHR}
efficiently. For an account of the most recent implementation see
\cite{Holzbaur:2001:OCC}.  Since, as already
mentioned above, many \texttt{CHR} rules are membership rules, our
approach provides a better implementation of a subset of \texttt{CHR}.
This, hopefully, may lead to new insights into a design and
implementation of languages appropriate for writing constraint
solvers.

The paper is organized as follows. In the next section we briefly
recall the original generic iteration algorithm of \cite{Apt00a}
and modify it successively to obtain the iteration algorithm
\texttt{R} for {\prop} rules. An important novelty is the preprocessing
phase during which we analyze the mutual dependencies between the
rules. This allows us to remove permanently some rules during the
iteration process. This permanent removal of the scheduled rules is
particularly beneficial in the context of constraint programming where
it leads to accumulated savings when constraint propagation is
intertwined with splitting.

In Section~\ref{sec:concrete} we recall the membership rules of
\cite{AM01} and show that they are {\prop} rules.
Then in Section~\ref{sec:implementation} we recall the relevant
aspects of the \texttt{CHR} language, discuss the implementation of
the \texttt{R} algorithm and present several benchmarks.
Finally, in Section~\ref{sec:redundancy}
we deal with the subject of redundancy of {\prop} rules.

\section{Revisions of the Generic Iteration Algorithm}
\label{sec:revised}

\subsection{The Original Algorithm}

Let us begin our presentation by recalling the generic algorithm of
\cite{Apt00a}. We slightly adjust the presentation to our purposes
by assuming that the considered partial ordering also has the
greatest element $\top$.

So we consider a partial ordering $(D, \po )$ with the
least element $\bot$ and the greatest element $\top$,
and a set of functions $F := \C{f_1, \LL , f_k}$
on $D$.  We are interested in functions that satisfy
the following two properties.
\begin{definition}
\begin{itemize}
\item $f$ is called {\em inflationary\/}
if $x \po f(x)$ for all $x$.

\item $f$ is called {\em monotonic\/}
if $x \po y$ implies
$f(x) \po f(y)$ for all $x, y$.
\HB
\end{itemize}
\end{definition}

The following algorithm is used to compute the least common
fixpoint of the functions from $F$.
\III

\NI
{\sc   Generic Iteration Algorithm ({\tt GI})}
\begin{tabbing}
\= $d := \bot$; \\
\> $G := F$; \\
\> {\bf while} $G \neq \ES$ {\bf and} $d \neq \top$ {\bf do} \\
\> \quad choose $g \in G$; \\
\> \quad $G := G - \C{g}$; \\
\> \quad $G := G \cup \update(G,g,d)$; \\
\> \quad $d := g(d)$ \\
\> {\bf end}
\end{tabbing}

where for all $G,g,d$ the set of functions
$\update(G,g,d)$ from $F$ is such that

\begin{description}
\item[{\bf A}] $\C{f \in F - G  \mid f(d) = d \A f(g(d)) \neq g(d)} \:\sse\: \update(G,g,d)$,

\item[{\bf B}] $g(d) = d$ implies that $\update(G,g,d) = \ES$,

\item[{\bf C}] $g(g(d)) \neq g(d)$ implies that $g \in \update(G,g,d)$.
\end{description}

Intuitively, assumption {\bf A} states that $\update(G,g,d)$ 
contains at least all the functions from $F - G$ for which the ``old value'',
$d$, is a fixpoint but the ``new value'', $g(d)$, is not.  So at each
loop iteration such functions are added to the set $G$.  In turn,
assumption {\bf B} states that no functions are added to $G$ in case
the value of $d$ did not change.  Assumption {\bf C} provides
information when $g$ is to be added back to $G$ as this information is
not provided by {\bf A}.
On the whole, the idea is to keep in $G$ at least all functions $f$
for which the current value of $d$ is not a fixpoint.

The use of the condition $d \neq \top$, absent in the original
presentation, allows us to leave the \textbf{while} loop earlier.
Our interest in the {\tt GI} algorithm is clarified by the
following result.

\begin{theorem}[Correctness] \label{thm:GI}
  Suppose that all functions in $F$ are inflationary and monotonic and
  that $(D, \po)$ is finite and has the least element $\bot$ and the
  greatest element $\top$.  Then every execution of the {\tt GI}
  algorithm terminates and computes in $d$ the least common fixpoint
  of the functions from $F$.
\end{theorem}
\Proof
Consider the predicate $I$ defined by:
\[
        I := (\fa f \in F - G \  f(d) = d) \A (\fa f \in F \ f(\top) = \top).
\]
Note that $I$ is established by the assignment $G := F$.
Moreover, it
is easy to check that by virtue of the assumptions {\bf A}, {\bf B} and {\bf C}
the predicate $I$ is preserved by each \textbf{while} loop iteration.  Thus $I$
is an invariant of the \textbf{while} loop of the algorithm.
Hence upon its termination
\[
        (G = \ES \Or d = \top) \A I
\]
holds, which implies
\[
        \fa f \in F \: f(d) = d.
\]

This implies that the algorithm computes in $d$ a common fixpoint
of the functions from $F$.

The rest of the proof is the same as in \cite{Apt00a}.
So the fact that this is the least common
fixpoint follows from the assumption that all functions are monotonic.

In turn, termination is established by considering
the lexicographic ordering of the strict partial orderings
$(D, \sqsupset)$ and $({\cal N}, <)$,
defined on the elements of $D \times {\cal N}$ by
\[
(d_1, n_1) <_{\mathit{lex}} (d_2, n_2) \quad\mbox{ iff }\quad d_1 \sqsupset d_2
        \ \mbox{ or }\ ( d_1 = d_2 \ \mbox{ and }\ n_1 < n_2).
\]
Then with each {\bf while} loop iteration of the algorithm the pair
$(d, card \: G)$, where $card \: G$ is the cardinality of the set $G$,
strictly decreases in the ordering $<_{\mathit{lex}}$.
\HB

\subsection{Removing Functions}

We now revise the {\tt GI} algorithm by modifying dynamically the set
of functions that are being scheduled. The idea is that, whenever
possible, we remove functions from the set $F$.  This will allow us to
exit the loop earlier which speeds up the execution of the algorithm.

To this end we assume that for each function $g \in F$ and each element $d \in D$,
two lists of functions from $F$ are given,
$\friends(g,d)$ and $\obviated(g,d)$,
to be instantiated below.
We then modify the {\tt GI} algorithm in such a way that each
application of $g$ to $d$ will be immediately followed by the
applications of all functions from $\friends(g,d)$ and by a removal of
the functions from $\friends(g,d)$ and from $\obviated(g,d)$ from
$F$ and $G$.  
Below we identify a condition, (\ref{eq:stable}), on $\friends(g,d)$ and
$\obviated(g,d)$ that ensures correctness of this scheduling strategy.
Informally, this condition states that after an application of all the functions 
from $\friends(g,d)$ the functions from $\friends(g,d)$ and from $\obviated(g,d)$
will not change anymore the subsequent values of $d$.

This modified algorithm has the following form.%
\footnote{We need in it lists instead of
sets since the considered functions will be applied in a specific order. But in
some places, for simplicity, we identify these lists with the sets.}

\III

\NI
{\sc Revised Generic Iteration Algorithm ({\tt RGI})}
\begin{tabbing}
\= $d := \bot$; \\
\> $F_0 := F$; \\
\> $G := F$; \\
\> {\bf while} $G \neq \ES$ {\bf and} $d \neq \top$ {\bf do} \\
\> \quad choose $g \in G$; \\
\> \quad $G := G - \C{g}$; \\
\> \quad $F := F - (\friends(g,d) \cup \obviated(g,d))$; \\
\> \quad $G := G - (\friends(g,d) \cup \obviated(g,d))$; \\
\> \quad $G$ \= $:= G \cup \update(G,h,d)$, \\
\>           \> where $\friends(g,d) = [g_1, \LL, g_k]$ and $h =  g \circ g_1 \circ \LL \circ g_k$;  \\
\> \quad $d :=h(d)$ \\
\> {\bf end}
\end{tabbing}

We now formalize the condition under which the
Correctness Theorem \ref{thm:GI} holds with the {\tt GI} algorithm
replaced by the {\tt RGI} algorithm.
To this end we consider the following property.
\begin{definition}
Suppose $d \in D$ and $f \in F$.
We say that $f$ is \emph{stable above $d$}
if $d \po e$ implies $f(e) = e$. We then say that $f$ is \emph{stable}
if it is stable above $f(d)$, for all $d$.
\HB
\end{definition}

That is, $f$ is stable if for all $d$ and $e$, $f(d) \po e$ implies
$f(e) = e$.  So stability implies idempotence, which means that
$f(f(d)) = f(d)$, for all $d$.
Moreover, if $d$ and $f(d)$ are comparable
for all $d$, then stability implies inflationarity.
Indeed, if d \po $f(d)$, then the claim holds vacuously.
And if $f(d) \po d$, then by stability $f(d) = d$.

Consider now the following condition
\begin{equation}
\fa d \: \fa e \sqsupseteq g \circ g_1 \circ \LL \circ g_k(d) \: \fa f \in \friends(g,d) \cup \obviated(g,d) \: (f(e) = e),
  \label{eq:stable}
\end{equation}
where $\friends(g,d) = [g_1, \LL, g_k]$.
That is, for all elements $d$, each function $f$ in
\mbox{$\friends(g,d) \cup \obviated(g,d)$}
is stable above $g \circ g_1 \circ \LL \circ g_k(d)$,
where $\friends(g,d)$ is the list $[g_1, \LL, g_k]$.
The following result holds.
\begin{theorem}
  Suppose that all functions in $F$ are inflationary and monotonic and
  that $(D, \po)$ is finite and has the least element $\bot$ and the
  greatest element $\top$.  Additionally, suppose that for each
  function $g \in F$ and $d \in D$ two lists of functions from $F$ are
  given, $\friends(g,d)$ and $\obviated(g,d)$ such that condition
  (\ref{eq:stable}) holds.

Then the Correctness Theorem \ref{thm:GI} holds with the {\tt GI}
algorithm replaced by the {\tt RGI} algorithm.
\end{theorem}
\Proof
In view of condition (\ref{eq:stable}) the following statement is an
invariant of the \textbf{while} loop:
\begin{equation}
        \fa f \in F - G \: (f(d) = d) \A
        \fa f \in F \ (f(\top) = \top) \A
        \fa f \in F_0 - F \ \fa e \sqsupseteq d \ (f(e) = e).
\label{eq:del}
\end{equation}
So upon termination of the algorithm the conjunction of this invariant with the negation of the loop condition, i.e.,
\[
        G = \ES \Or d = \top
\]
holds, which implies that $\fa f \in F_0 \: (f(d) = d)$.

The rest of the proof is the same.
\HB

\subsection{Functions in the Form of Rules}

In what follows we consider the situation when the scheduled functions
are of a specific form $b \ra g$, where $b$ is a \emph{condition} and
$g$ a function, that we call a \emph{body}. We call such functions \emph{rules}.

First, we explain how rules are applied.
Given an element $d$ of $D$ a condition $b$ is evaluated in $d$.
The outcome is either \emph{true}, that we denote by $\holds bd$, or \emph{false}.

Given a rule $b \ra g$ we define then its application to $d$
as follows:
\[
        (b \ra g)(d) :=
        \left\{
        \begin{array}{l@{\extracolsep{3mm}}l}
        g(d)    & \mathrm{if}\ \holds bd\\
        d       & \mathrm{otherwise}
        \end{array}
        \right.
\]

We are interested in a specific type of conditions and rules.
\begin{definition}
Consider a partial ordering $(D, \po )$.
\begin{itemize}

\item We say that a condition $b$ is \emph{monotonic} if
for all $d, e$ we have that
$\holds bd$ and $d \po e$ implies $\holds be$.

\item We say that a condition $b$ is \emph{precise} if
the least $d$ exists such that
$\holds bd$.
We call then $d$ the \emph{witness} for $b$.

\item  We call a rule $b \ra g$ a \emph{{\prop}} rule
if $b$ is monotonic and precise and $g$ is stable.
\hspace*{3em}\HB
\end{itemize}
\end{definition}

To see how natural this class of rules is consider the following example.

\begin{example}
Take a set $A$ and consider the partial ordering
\[
        ({\cal P}(A), \subseteq).
\]
In this ordering the empty set $\ES$ is the least element and $A$ is the greatest
element.
We consider rules of the form
\[
        B \ra G,
\]
where $B, G \sse A$.

To clarify how they are applied to subsets of $A$ we first stipulate for $E \sse A$
\[
        \holds BE \mbox{ iff } B \sse E.
\]
Then we view a set $G$ as a function on ${\cal P}(A)$ by putting
\[
        G(E) := G \cup E.
\]
This determines the rule application of $B \ra G$.

It is straightforward to see that such rules are {\prop} rules.
In particular,
the element $B$ of ${\cal P}(A)$ is the witness for the condition $B$.
For the stability of $G$ take $E \sse A$ and suppose $G(E) \sse F$.
Then $G(E) = G \cup E$, so $G \cup E \sse F$,
which implies $G \cup F = F$, i.e., $G(F) = F$.

Such rules can be instantiated to many situations.  For example, we
can view the elements of the set $A$ as primitive constraints. Then
each rule $B \ra G$ is a natural operation on the constraint store: if
all constraints in $B$ are present in the store, then add to it all
constraints in $G$.

Alternatively, we can view $A$ as a set of some atomic formulas
and each rule $B \ra G$ as a proof rule, usually written as
\[
        \frac{B}{G}
\]
The minor difference with the usual proof-theoretic framework is that
rules have then a single conclusion.
An axiom is then a rule with the empty set $\ES$ as the condition.
A closure under such a set of rules is then the set of all
(atomic) theorems that can be proved using them.

The algorithm presented below can in particular be used
to compute efficiently the closure under such proof rules
given a finite set of atomic formulas $A$.
\HB
\end{example}

We now modify the \texttt{RGI} algorithm for the case of {\prop} rules.
In the algorithm below we take into account that an application of a
rule is a two step process: testing of the condition followed by a
conditional application of the body.
This will allow us to drop the parameter $d$
from the lists $\friends(g,d)$ and $\obviated(g,d)$
and consequently to construct such lists before the execution
of the algorithm.
The list $\friends(g)$ will be constructed in such a way that we shall
not need to evaluate the conditions of its rules: they will all hold.
The specific construction of the lists $\friends(g)$ and $\obviated(g)$
that we use here will be
provided in the second algorithm, called {\sc Friends and Obviated Algorithm}.
\newpage

\III

\NI
{\sc Rules Algorithm ({\tt R})}
\begin{tabbing}
\= $d := \bot$; \\
\> $F_0 := F$; \\
\> $G := F$; \\
\> {\bf while} $G \neq \ES$ {\bf and} $d \neq \top$ {\bf do} \\
\> \quad choose $f \in G$; suppose $f$ is $b \ra g$; \\
\> \quad $G := G - \C{b \ra g}$; \\
\> \quad {\bf if}\  $\holds bd$ {\bf then} \\
\> \quad \quad $F := F - (\friends(b \ra g) \cup \obviated(b \ra g))$; \\
\> \quad \quad $G := G - (\friends(b \ra g) \cup \obviated(b \ra g))$; \\
\> \quad \quad $G$ \= $:= G \cup \update(G,h,d)$, \\
\>                 \> where $\friends(b \ra g)= [b_1 \ra g_1, \LL, b_k \ra g_k]$ and
$h = g \circ g_1 \circ \LL \circ g_k$;  \\
\> \quad \quad $d :=h(d)$ \\
\> \quad {\bf else if}\  $\forall e \sqsupseteq d \; \neg\holds be$ {\bf then}\\
\> \quad \quad \quad $F := F - \{ b \ra g \}$ \\
\> \quad {\bf end} \\
\> {\bf end}
\end{tabbing}

Again, we are interested in identifying conditions under which the
Correctness Theorem \ref{thm:GI} holds with the {\tt GI} algorithm
replaced by the {\tt R} algorithm.  To this end, given a rule $b \ra g$
in $F$ and $d \in D$ define as follows:
\[
\begin{array}{lrcl}&
\friends(b \ra g, d)  & := &
\left\{
\begin{array}{ll}
\friends(b \ra g) &  \mbox{if } \holds bd,\\{}
[ \ ]             &  \mbox{otherwise}
\end{array}
\right.
\\\hspace*{-2.3em}\mbox{and}\\&
\obviated(b \ra g, d)  & := &
\left\{
\begin{array}{ll}
\obviated(b \ra g) &  \mbox{if } \holds bd,\\{}
[ b \ra g ]        &  \mbox{if } \forall e \sqsupseteq d \; \neg\holds be,\\{}
[ \ ]              &  \mbox{otherwise}
\end{array}
\right.
\end{array}
\]

We obtain the following counterpart of the Correctness Theorem \ref{thm:GI}.
\begin{theorem}[Correctness] \label{thm:R}
  Suppose that all functions in $F$ are {\prop} rules of the form $b \ra
  g$, where $g$ is inflationary and monotonic, and that $(D, \po)$ is
  finite and has the least element $\bot$ and the greatest element
  $\top$.  Further, assume that for each rule $b \ra g$ the lists
  $\friends(b \ra g,d)$ and $\obviated(b \ra g,d)$ defined as above
  satisfy condition~(\ref{eq:stable}) and the following condition:
\begin{equation}
\fa d (\; b' \ra g' \in \friends(b \ra g) \A \holds bd
        \;\ra\;
        \fa e \sqsupseteq g(d) \ \holds {b'}e \;).
\label{eq:friends}
\end{equation}
Then the Correctness Theorem \ref{thm:GI} holds with the {\tt GI}
algorithm replaced by the {\tt R}~algorithm.
\end{theorem}
\Proof
It suffices to show that the {\tt R} algorithm is an instance of the
{\tt RGI} algorithm.
On the account of condition (\ref{eq:friends}) and the fact that the
rule bodies are inflationary functions, $\holds bd$
implies that
\[
        ((b \ra g) \circ (b_1 \ra g_1) \circ \LL \circ (b_k \ra g_k))(d)
        \ =\  (g \circ g_1 \circ \LL \circ g_k)(d),
\]
where $\friends(b \ra g) = [b_1 \ra g_1, \LL, b_k \ra g_k]$.
This takes care of the situation in which $\holds bd$ is true.

In turn, the definition of $\friends(b \ra g,d)$ and $\obviated(b \ra g,d)$
and assumption \textbf{B} take care of the situation when $\neg\holds bd$.
When the condition $b$ fails for all $e \sqsupseteq d$ we can
conclude that for all such $e$ we have $(b \ra g)(e)=e$.  This allows
us to remove at that point of the execution the rule $b \ra g$ from
the set $F$.  This amounts to adding $b \ra g$ to the set
$\obviated(b \ra g, d)$ at runtime. Note that
condition~(\ref{eq:stable}) is then satisfied.
\HB

We now provide an explicit construction of the lists $\friends$ and
$\obviated$ for a rule $b \ra g$ in the form of the following
algorithm.  $\mathtt{GI}(d)$ stands here for the \texttt{GI}
algorithm activated with $\bot$ replaced by $d$ and the considered set
of rules as the set of functions $F$.
Further, given an execution of $\mathtt{GI}(e)$,
we call here a rule $g$ \emph{relevant} if at some point $g(d) \neq d$ holds
after the ``choose $g \in G$'' action.
\III

\NI
{\sc Friends and Obviated Algorithm ({\tt F \& O})}
\begin{tabbing}
\= $w :=$ witness of $b$;\\
\> $d := \mathtt{GI}(g(w))$; \\
\> $\friends(b \ra g) :=$ \=list of the relevant rules $h \in F$ in the execution of $\mathtt{GI}(g(w))$;\\
\> $\obviated(b \ra g) := [\ ]$; \\
\> {\bf for each} $(b' \ra g') \in F - \friends(b \ra g)$ {\bf do} \\
\> \quad {\bf if}  $g'(d) = d$\ {\bf or}\  $\fa e \sqsupseteq d \ \neg\holds {b'}{e}$ {\bf then} \\
\> \quad \quad $\obviated(b \ra g) := [\;b' \ra g' \;|\; \obviated(b \ra g)\;]$ \\
\> \quad {\bf end} \\
\> {\bf end}
\end{tabbing}
Note that $b \ra g$ itself is not contained in $\friends(b \ra g)$
as it is a \emph{prop} rule,
however it is in $\obviated(b \ra g)$,
since by the stability of $g$ \ $g(d) = d$ holds.

The following observation now shows the adequacy of the {\tt F \& O} algorithm
for our purposes.

\begin{lemma}
Upon termination of the {\tt F \& O} algorithm
conditions (\ref{eq:stable}) and (\ref{eq:friends}) hold,
where the lists $\friends(b \ra g,d)$ and $\obviated(b \ra g,d)$
are defined as before Theorem \ref{thm:R}.
\HB
\end{lemma}

Let us summarize now the findings of this section that culminated in
the {\tt R} algorithm.  Assume that all functions are of the form of
the rules satisfying the conditions of the Correctness Theorem
\ref{thm:R}. Then in the {\tt R} algorithm, each time the evaluation
of the condition $b$ of the selected rule $b \ra g$  succeeds,

\begin{itemize}
\item the rules in the list $\friends(b \ra g)$ are applied
directly without testing the value of their conditions,

\item the rules in $\friends(b \ra g) \cup \obviated(b \ra g)$
are permanently removed from the current set of functions $G$
and from $F$.
\end{itemize}

\subsection{Recomputing of the Least Fixpoints}

Another important optimization takes place when the {\tt R} algorithm is
repeatedly applied to compute the least fixpoint.  More specifically,
consider the following sequence of actions:
\begin{itemize}
\item we compute the least common fixpoint $d$ of the functions from $F$,

\item we move from $d$ to an element $e$ such that $d \po e$,

\item we compute the least common fixpoint above $e$ of
the functions from $F$.
\end{itemize}
Such a sequence of actions typically arises in the framework of CSPs,
further studied in Section~\ref{sec:concrete}.
The computation of the least common fixpoint $d$
of the functions from $F$ corresponds there to the constraint
propagation process for a constraint $C$.  The moving from $d$ to
$e$ such that $d \po e$ corresponds to splitting or constraint
propagation involving another constraint, and the computation of the
least common fixpoint above $e$ of the functions from $F$ corresponds
to another round of constraint propagation for $C$.

Suppose now that we computed the least common fixpoint $d$ of the
functions from $F$ using the {\tt RGI} algorithm or its modification
\texttt{R} for the rules. During its execution we permanently removed
some functions from the set $F$.  These functions are then not
needed for computing the least common fixpoint above $e$
of the functions from $F$.  The precise statement is provided in
the following simple, yet crucial, theorem.

\begin{theorem} \label{thm:repeated}
  Suppose that all functions in $F$ are inflationary and monotonic and
  that $(D, \po)$ is finite. Suppose that the least common fixpoint
  $d_0$ of the functions from $F$ is computed by means of the {\tt RGI}
  algorithm or the {\tt R} algorithm.  Let $F_{\mathit{fin}}$ be the
  final value of the variable $F$ upon termination of the {\tt RGI}
  algorithm or of the {\tt R} algorithm.

  Suppose now that $d_0 \po e$. Then the least common fixpoint $e_0$ above $e$
  of the functions from $F$ coincides with the least common fixpoint
  above $e$ of the functions from $F_{\mathit{fin}}$.
\end{theorem}
\Proof
Take a common fixpoint $e_1$ of the functions from $F_{\mathit{fin}}$
such that $e \po e_1$.  It suffices to prove that $e_1$ is common
fixpoint of the functions from $F$.  So take $f \in F - F_{\mathit{fin}}$.
Since condition (\ref{eq:del}) is an invariant of the main
\textbf{while} loop of the {\tt RGI} algorithm and of the {\tt R}
algorithm, it holds upon termination and consequently $f$ is stable
above $d_0$.  But $d_0 \po e$ and $e \po e_1$, so we conclude that
$f(e_1) = e_1$.
\HB
\smallskip

Intuitively, this result means that if after splitting we relaunch the
same constraint propagation process we can disregard the removed
functions.

In the next section we instantiate the \texttt{R} algorithm by a set
of rules that naturally arise in the context of constraint
satisfaction problems with finite domains. In Section~\ref{sec:implementation}
we assess the practical impact of the discussed optimizations.

\section{Concrete Framework}
\label{sec:concrete}

We now proceed with the main topic of this paper, the schedulers for
the rules that naturally arise in the context of constraint
satisfaction problems.  First we recall briefly the necessary
background on the constraint satisfaction problems.

\subsection{Constraint Satisfaction Problems}

Consider a sequence of variables $X := x_1, \LL, x_n$
where $n \geq 0$, with respective domains $D_1, \LL, D_n$
associated with them.  So each variable $x_i$ ranges over the domain
$D_i$.  By a {\em constraint} $C$ on $X$ we mean a subset of $D_1
\times \LL \times D_n$.
Given an element $d := d_1, \LL, d_n$ of $D_1 \times \LL \times D_n$
and a subsequence $Y := x_{i_1}, \LL, x_{i_\ell}$ of $X$ we denote by
$d[Y]$ the sequence $d_{i_1}, \LL, d_{i_{\ell}}$. In particular, for a
variable $x_i$ from $X$, $d[x_i]$ denotes $d_i$.

Recall that a {\em constraint satisfaction problem}, in short CSP, consists of
a finite sequence of variables $X$ with respective domains ${\cal D}$,
together with a finite set $\cal C$ of constraints, each on a
subsequence of $X$. We write it as
$\p{{\cal C}}{x_1 \in D_1, \LL, x_n \in D_n}$,
where $X := x_1, \LL, x_n$ and ${\cal D} := D_1, \LL, D_n$.

By a {\em solution\/} to $\p{{\cal C}}{x_1 \in D_1, \LL, x_n \in D_n}$
we mean an element $d \in D_1 \times \LL \times D_n$ such that for
each constraint $C \in {\cal C}$ on a sequence of variables $X$ we
have $d[X] \in C$.  We call a CSP {\em consistent\/} if it has a
solution.  Two CSPs with the same sequence of variables are called
{\em equivalent\/} if they have the same set of solutions.

\subsection{Partial Orderings}

With each CSP ${\cal P} := \p{{\cal C}}{x_1 \in D_1, \LL, x_n \in D_n}$
we associate now a specific partial ordering.  Initially we
take the Cartesian product of the partial orderings $({\cal P}(D_1),
\supseteq ), \LL, ({\cal P}(D_n), \supseteq)$.
So this ordering is of the form
\[
        ({\cal P}(D_1) \times  \LL \times {\cal P}(D_n), \supseteq)
\]
where we interpret $\supseteq$ as
the Cartesian product of the reversed subset ordering.
The elements of this partial ordering are sequences
$(E_1, \LL, E_n)$ of respective subsets of $(D_1, \LL, D_n)$ ordered
by the component-wise reversed subset ordering. Note that in this
ordering $(D_1, \LL, D_n)$ is the least element while
\[
        \underbrace{(\ES, \LL, \ES)}_{n\ \mathrm{times}}
\]
is the greatest element. However, we would like to identify with the
greatest element all sequences that contain as an element the empty
set.  So we divide the above partial ordering by the equivalence
relation $R_{\ES}$ according to which

\begin{tabbing}
\qquad \qquad \qquad  $(E_1, \LL, E_n)  \ R_{\ES} \ (F_1, \LL, F_n)$ iff \= $(E_1, \LL, E_n) = (F_1, \LL, F_n)$\\
        \> or ($\te i \: E_i = \ES$ and $\te j \: F_j = \ES$).
\end{tabbing}
It is straightforward to see that $R_{\ES}$ is indeed an equivalence relation.

In the resulting quotient ordering there are two types of elements: the
sequences $(E_1, \LL, E_n)$ that do not contain the empty set as an
element, that we continue to present in the usual way with the
understanding that now each of the listed sets is non-empty, and one
``special'' element equal to the equivalence class consisting of all
sequences that contain the empty set as an element. This equivalence
class is the greatest element in the resulting ordering, so we denote
it by $\top$.  In what follows we denote this partial ordering by
$(D_{\cal P}, \po )$.

\subsection{Membership Rules}

Fix now a specific CSP ${\cal P} := \p{{\cal C}}{x_1 \in D_1, \LL, x_n \in D_n}$
with finite domains. We recall the
rules introduced in \cite{AM01}.
They are called {\em membership rules} and are of the form
\[
        y_1 \in S_1, \LL, y_k \in S_k \;\ra\; z_1 \neq a_1, \LL, z_m \neq a_m,
\]
where

\begin{itemize}
\item $y_1, \LL, y_k$ are pairwise different variables from the set $\C{x_1, \LL, x_n}$
  and $S_1, \LL, S_k$ are subsets of the respective variable domains,

\item  $z_1, \LL, z_m$ are variables from the set $\C{x_1, \LL, x_n}$
and $a_1, \LL, a_m$ are elements of the respective variable domains.%
\footnote{In \cite{AM01} it is also assumed that
the lists $y_1, \LL, y_k$ and $z_1, \LL, z_m$
have no variable in common. We drop this condition so that we can combine
the membership rules.}
\end{itemize}
Note that we do not assume that the variables $z_1, \LL, z_m$
are pairwise different.

The computational interpretation of such a rule is:

\begin{quote}
if for $i \in [1..k]$ the current domain of the variable $y_i$
is included in the set $S_i$, then
for $j \in [1..m]$ remove the element $a_i$ from the
domain of $z_i$.
\end{quote}
When each set $S_i$ is a singleton, we call a membership rule an \emph{equality rule}.

Let us mention here that in \cite{AM01} the interpretation of the conditions
of an equality rule is slightly different, as it is stipulated that
the current domain of the variable $y_i$ is to be equal to the singleton set $S_i$.
However, in the discussed algorithms
the membership rules are applied only when all variable domains are non-empty
and then both interpretations coincide.

Let us reformulate this interpretation so that it fits the framework
considered in the previous section.  To this end we need to clarify how to
\begin{itemize}
\item evaluate the condition of a membership rule in an element
of the considered partial ordering,
\item interpret the conclusion of a membership rule
as a function on the  considered partial ordering.
\end{itemize}
Let us start with the first item.
\begin{definition}
Given a variable $y$ with the domain $D_y$ and $E \in {\cal P}(D_y)$ we define
\[
        \holds {y \in S}E \quad\text{iff}\quad E \sse S,
\]
and extend the definition to the elements of the considered ordering
$(D_{\cal P}, \po )$ by putting
\[\begin{array}{l}
        \holds {y \in S}{(E_1, \LL, E_n)} \quad\text{iff}\quad E_k \sse S,
        \quad\text{where we assumed that $y$ is $x_k$, and}
\\
        \holds {y \in S}{\top}.
\end{array}\]

Furthermore we interpret a sequence of conditions as a conjunction, by putting
\[\begin{array}{l}
\holds {(y_1 \in S_1, \LL, y_k \in S_k)}{\,(E_1, \LL, E_n)}
\\\quad\text{iff}\quad
\holds {y_i \in S_i}{\,(E_1, \LL, E_n)} \quad\text{for}\ i \in [1..k].
\end{array}\]
\HB
\end{definition}

Concerning the second item we proceed as follows.
\begin{definition}
  Given a variable $z$ with the domain $D_z$ we interpret the
  atomic formula $z \neq a$ as a function on ${\cal P}(D_z)$,
  defined by:
\[
        (z \neq a)(E) := E - \C{a}.
\]

Then we extend this function to the elements of the considered
ordering $(D_{\cal P}, \po )$ as follows:
\begin{itemize}
\item on the elements of the form $(E_1, \LL, E_n)$
we put
\[
        (z \neq a)(E_1, \LL, E_n) := (E'_1, \LL, E'_n),
\]
where
\begin{itemize}
\item if $z \equiv x_i$, then $E'_i = E_i - \C{a}$,
\item if $z \not \equiv x_i$, then $E'_i = E_i$.
\end{itemize}
If the resulting sequence $(E'_1, \LL, E'_n)$ contains the empty set,
we replace it by~$\top$,

\item on the element $\top$ we put
$(z \neq a)(\top) := \top$
\end{itemize}
Finally, we interpret a sequence $z_1 \neq a_1, \LL, z_m \neq a_m$
of atomic formulas by interpreting each of them in turn.
\HB
\end{definition}

As an example take the CSP
\[
        {\cal P} := \p{{\cal C}}{x_1 \in \C{a,b,c}, x_2 \in \C{a,b,c}, x_3 \in \C{a,b,c}, x_4 \in \C{a,b,c}}
\]
and consider the membership rule
\[
        r := \ \  x_1 \in \C{a,b}, x_2 \in \C{b} \;\ra\; x_3 \neq a, x_3 \neq b, x_4 \neq a.
\]
Then we have

\[
\begin{array}{rcl}
        r(\C{a}, \C{b}, \C{a,b,c}, \C{a,b}) &=& (\C{a}, \C{b}, \C{c}, \C{b}),\\
        r(\C{a,b,c}, \C{b}, \C{a,b,c}, \C{a,b}) &=& (\C{a,b,c}, \C{b}, \C{a,b,c}, \C{a,b}),\\
        r(\C{a,b}, \C{b}, \C{a,b}, \C{a,b}) &=& \top.
\end{array}
\]

In view of the Correctness Theorem \ref{thm:R} the following
observation allows us to apply the \texttt{R} algorithm when each
function is a membership rule and when for each rule
$b \ra g$ the lists $\friends(b \ra g)$ and $\obviated(b \ra g)$ are
constructed by the {\tt F~\&~O} algorithm.

\begin{note}
Consider the partial ordering $(D_{\cal P}, \po )$.

\begin{enumerate}\smallromani
\item Each membership rule is a {\prop} rule.

\item 
Each function $z_1 \neq a_1, \LL, z_m \neq a_m$ on $D_{\cal P}$ is
\begin{itemize}
\item inflationary,

\item monotonic.
\end{itemize}
\vspace{-3ex}
\end{enumerate}
\HB
\end{note}

To be able to instantiate the algorithm \texttt{R}
with the membership rules we still need to define the set
$\update(G,g,d)$. In our implementation we chose the following
simple definition:
\[
        \update(G,b \ra g,d) :=
        \left\{
        \begin{array}{ll}
        F-G & \mbox{if } \holds bd \mbox{ and } g(d) \neq d, \\
        \ES & \mbox{otherwise.}
        \end{array}
        \right.
\]

To illustrate the intuition behind the use of the lists $\friends(b \ra g)$ and $\obviated(b \ra g)$
take the above CSP
\[
        {\cal P} := \p{{\cal C}}{x_1 \in \C{a,b,c}, x_2 \in \C{a,b,c}, x_3 \in \C{a,b,c}, x_4 \in \C{a,b,c}}
\]
and consider the membership rules
\[
\begin{array}{llcl}
        r_1 := \ & x_1 \in \C{a,b} &\ra& x_2 \neq a, x_4 \neq b,\\
        r_2 := & x_1 \in \C{a,b}, x_2 \in \C{b,c} &\ra& x_3 \neq a,\\
        r_3 := & \hspace{5.1em} x_2 \in \C{b} &\ra& x_3 \neq a, x_4 \neq b.\\
\end{array}
\]
Then upon application of rule $r_1$ rule $r_2$ can
be applied without evaluating its condition and subsequently
rule $r_3$ can be deleted without applying it.
So we can put rule $r_2$ into $\friends(r_1)$ and rule $r_3$ into $\obviated(r_1)$
and this is in fact what the {\tt F \& O} algorithm does.

\section{Implementation}
\label{sec:implementation}

In this section we discuss the implementation of the \texttt{R}
algorithm for the membership rules and compare it by means of various
benchmarks with the \texttt{CHR} implementation in the \eclipse{} system.

\subsection{Modelling of the Membership Rules in \texttt{CHR}}

Following \cite{AM01} the membership rules are
represented as \texttt{CHR} propagation rules with one head.  Recall
that the latter ones are of the form
\[
        H ==> G_1, \ldots, G_l ~|~ B_1, \ldots, B_m.
\]
where
\begin{itemize}
\item $l \geq 0$, $m > 0$,

\item the atom $H$ of the \emph{head} refers to
the defined constraints,

\item the atoms of the \emph{guard} $G_1, \ldots,
G_l$ refer to Prolog relations or built-in constraints,

\item the atoms of the \emph{body} $B_1, \ldots, B_m$ are arbitrary atoms.
\end{itemize}

Further, recall that the {\tt CHR} propagation rules with one head are
executed as follows.  First, given a query (that represents a CSP) the
variables of the rule are renamed to avoid variable clashes.  Then an
attempt is made to match the head of the rule against the first atom
of the query.  If it is successful and all guards of the instantiated version
of the rule succeed, the instantiated version of the body of the rule is
executed. Otherwise the next rule is tried.

Finally, let us recall the representation of a membership rule as
\texttt{CHR} a propagation rule used in \cite{AM01}.
Consider the membership rule
\[
        y_1 \in S_1, \LL, y_k \in S_k \;\ra\; z_1 \neq a_1, \LL, z_m \neq a_m.
\]
related to the constraint \texttt{c} on the variables $X_1, \LL, X_n$.
We represent its condition by starting initially with the atom $c(X_1,
\LL, X_n)$ as the head.  Each atomic condition of the form $y_i \in
\C{a}$ is processed by replacing in the atom $c(X_1, \LL, X_n)$ the
variable $y_i$ by $a$. In turn, each atomic condition of the form $y_i
\in S_i$, where $S_i$ is not a singleton, is processed by adding the
atom $\mathtt{in(y_i, LS_i)}$ to the guard of the propagation rule.
The {\tt in/2} predicate is defined by
\[
        \verb+in(X,L) :- dom(X,D), subset(D,L).+
\]
So {\tt in(X,L)} holds if the current domain of the variable {\tt
X} (yielded by the built-in {\tt dom} of \eclipse{}) is included in
the list {\tt L}.
In turn, $\mathtt{LS_i}$ is a list representation of the set $S_i$.

Finally, each atomic conclusion $z_i \neq a_i$ translates to the atom
$z_i$ \verb?##? $a_i$ of the body of the propagation rule.

As an example consider the membership rule
\[
        X \in \C{0}, Y \in \C{1,2} \;\ra\; Z \neq 2
\]
in presence of a constraint $c$ on the variables $X,Y,Z$.
It is represented by the following {\tt CHR} propagation rule:
\[
        \verb+c(0,Y,Z) ==> in(Y,[1,2]) | Z##2.+
\]

In \eclipse{} the variables with singleton domains are automatically
instantiated. So, assuming that the variable domains are non-empty,
the condition of this membership rule holds iff the head of the
renamed version of the above propagation rule matches the atom
\verb+c(0,Y,Z)+ \emph{and} the current domain of the variable
\verb+Y+ is included in \verb+[1,2]+.  Further, in both cases the
execution of the body leads to the removal of the value 2 from the
domain of \verb+Z+.  So the execution of both rules has the same
effect when the variable domains are non-empty.

\paragraph{Execution of CHR.}
In general, the application of a membership rule as defined in
Section~\ref{sec:concrete} and the execution of its representation
as a {\tt CHR} propagation rules coincide.
Moreover, by the semantics of
\texttt{CHR}, the \texttt{CHR} rules are repeatedly applied until a
fixpoint is reached.  So a repeated
application of a finite set of membership rules coincides with the
execution of the \texttt{CHR} program formed by the representations of
these membership rules as propagation rules.
An important point concerning the standard execution
of a \texttt{CHR} program is that, in contrast
to the \texttt{R} algorithm, every change in the variable domains
of a constraint causes the computation to restart.

\subsection{Benchmarks}

In our approach the repeated application of a finite set of
membership rules is realized by means of the \texttt{R} algorithm of
Section~\ref{sec:revised} implemented in \eclipse.
The compiler consists of about 1500 lines of code.
It accepts as input a set of membership rules, each represented as a
\texttt{CHR} propagation rule, and constructs an \eclipse{} program
that is the instantiation of the \texttt{R} algorithm for this set of
rules.  As in \texttt{CHR}, for each constraint the set of rules that
refer to it is scheduled separately.

In the benchmarks below for each considered CSP we used the sets of
all minimal valid membership and equality rules for the ``base''
constraints which were automatically generated using a program
discussed in \cite{AM01}.  In the first phase the
compiler constructs for each rule $g$ the lists $\friends(g)$ and
$\obviated(g)$.  Time spent on this construction is comparable with
the time needed for the generation of the minimal valid equality and
membership rules for a given constraint.
For example, the medium-sized membership rule set for the \texttt{rcc8} constraint,
containing 912 rules, was generated in $166$~seconds while the construction of all 
$\friends$ and $obviated$ lists took $142$ seconds.

To see the impact of the accumulated savings obtained by permanent
removal of the rules during the iteration process, we chose
benchmarks that embody several successive propagation steps,
i.e., propagation interleaved with domain splitting or labelling.

In Table \ref{tab:fixpoints} we list the results for selected single
constraints.  For each such constraint, say $C$ on a sequence of
variables $x_1, \LL, x_n$ with respective domains $D_1, \LL, D_n$, we
consider the CSP $\p{C}{x_1 \in D_1, \LL, x_n \in D_n}$ together with
randomized labelling. That is, the choices of variable, value, and action
(assigning or removing the value), are all random.  The computation of simply
one or all solutions yields insignificant times, so the benchmark
program computes and records also all intermediate non-solution fixpoints.
Backtracking occurs if a recorded fixpoint is encountered again.
In essence, this benchmark computes implicitly all possible search trees.
As this takes too much time for some constraints, we also impose a limit on the
number of recorded fixpoints.

In turn, in Table \ref{tab:atpg} we list the results for selected
CSPs. We chose here CSPs that formalize
sequential automatic test pattern generation for digital circuits
(ATPG), see \cite{brand:atpg:2001}.
These are rather large CSPs that employ the \texttt{and}
constraints of Table \ref{tab:fixpoints} and a number of other
constraints, most of which are implemented by rules.

We measured the execution times for three rule schedulers:
the standard \texttt{CHR} representation of the rules,
the generic chaotic iteration algorithm \texttt{GI},
and its improved derivative \texttt{R}.
The codes of the latter two algorithms are both produced by
our compiler and are structurally equal, hence allow a direct
assessment of the improvements embodied in~\texttt{R}.

\begin{table}[htbp]
  \begin{center}
\begin{tabular*}{\textwidth}{l@{\extracolsep{\fill}}ccccc}\hline\hline\\[-2ex]
{\bfseries Constraint}\hspace*{-0.3em}
        & \texttt{rcc8}
        & \texttt{fork}
        & \texttt{and3}
        & \texttt{and9}
        & \texttt{and11}
        \\\hline
\textsc{membership}\hspace*{-0.3em}\\[1ex]
\quad relative
        & 37\% / 22\%
        & 58\% / 46\%
        & 66\% / 49\%
        & 26\% / 15\%
        & 57\% / 25\%
        \\[1mm]
\quad absolute
        & {\small 147/396/686}
        & {\small 0.36/0.62/0.78}
        & {\small 0.27/0.41/0.55}
        & {\small 449/1727/2940}
        & \hspace*{-0.2em}{\small 1874/3321/7615}
        \\[2mm]
\textsc{equality}\\[1ex]
\quad relative
        & 97\% / 100\%
        & 98\% / 94\%
        & 92\% / 59\%
        & 95\% / 100\%
        & 96\% / 101\%
        \\[1mm]
\quad absolute
        & {\small 359/368/359}
        & {\small 21.6/21.9/22.9}
        & {\small 0.36/0.39/0.61}
        & {\small 386/407/385}
        & {\small 342/355/338}
\\\hline\hline\\[-1ex]
\end{tabular*}
    \caption{Randomized search trees for single constraints}
    \label{tab:fixpoints}
  \end{center}
\end{table}

\begin{table}[htbp]
  \begin{center}
\begin{tabular*}{\textwidth}{l@{\extracolsep{\fill}}cccc}\hline\hline\\[-2ex]
{\bfseries Logic}
        & 3-valued
        & 9-valued
        & 11-valued
        \\\hline
\textsc{membership}\\[1ex]
\quad relative
        & 61\% / 44\%
        & 65\% / 29\%
        & 73\% / 29\%
        \\[1mm]
\quad absolute
        & {\small 1.37/2.23/3.09}
        & {\small 111/172/385}
        & {\small 713/982/2495}
        \\[2mm]
\textsc{equality}\\[1ex]
\quad relative
        & 63\% / 29\%
        & 40\% / 57\%
        & 36\% / 51\%
        \\[1mm]
\quad absolute
        & {\small 0.77/1.22/2.70}
        & {\small 2.56/6.39/4.50}
        & {\small 13.8/38.7/26.7}
\\\hline\hline\\[-1ex]
\end{tabular*}
    \caption{CSPs formalizing sequential ATPG}
    \label{tab:atpg}
  \end{center}
\end{table}

An important point in the implementations is the question of when to
remove solved constraints from the constraint store.  The standard
\texttt{CHR} representation of membership rules as generated by the
algorithm of \cite{AM01} does so by containing, beside the propagation
rules, one \texttt{CHR} simplification rule for each tuple in
the constraint definition.  Once its variables are
assigned values that correspond to a tuple, the constraint is solved,
and removed from the store by the corresponding simplification rule.
This `solved' test takes place interleaved with propagation.
The implementations of \texttt{GI} and \texttt{R}, on the other hand,
check after closure under the propagation rules.  The constraint is
considered solved if all its variables are fixed, or, in the case of
\texttt{R}, if the set $F$ of remaining rules is empty (see the
following subsection).  Interestingly, comparing \texttt{CHR} and
\texttt{GI}, the extra simplification rules sometimes constitute a
substantial overhead while at other times their presence allows
earlier termination.

We mention briefly that our specific implementation deviates slightly
from the description of \texttt{R} inside the \texttt{else} branch.
The test $\forall e \sqsupseteq d \; \neg\holds be$
in the case of a membership condition $y \in S$ corresponds to testing
whether the intersection $D_y \cap S$ is empty.  Performing this always
turned out to be more costly than doing so only when $D_y$ is a
singleton set.

The platform for all benchmarks was a Sun Enterprise 450
with 4 UltraSPARC-II $400\,$MHz processors and $2\,$GB memory
under Solaris, and {\eclipse} 5.5 (non-parallel).
In the tables we provide for each constraint or CSP the ratio of
the execution times in seconds between,
first, \texttt{R} and \texttt{GI}, and second,
\texttt{R} and \texttt{CHR}.  This is followed by the
absolute times in the order \texttt{R} / \texttt{GI} / \texttt{CHR}.

Recently, we have been experimenting with various ways of optimizing
our implementation of the \texttt{R} algorithm.  In particular,
we considered a better embedding into the constraint-handling mechanism
of \eclipse, for instance by finer control of the waking conditions
and a joint removal of the elements from the same domain.
At this stage we succeeded in achieving an average speed-up
by a factor of $4$.  This work is in progress but already shows
that further improvements are possible.

\subsection{Recomputing of the Least Fixpoints}

\newcommand{\kleene}[1]{\mathsf{#1}}
\newcommand{\uuu}{\kleene{u}}
\newcommand{\ttt}{\kleene{t}}
\newcommand{\fff}{\kleene{f}}

Finally, let us illustrate the impact of the permanent removal of the
rules during the least fixpoint computation, achieved here by the use
of the lists $\friends(g)$ and $\obviated(g)$.  Given a set $F$ of
rules call a rule $g \in F$ \emph{solving} if
\mbox{$\friends(g) \cup \obviated(g) = F$}.

Take now as an example the equivalence relation $\equiv$ from three valued logic of
\cite[page 334]{Kle52} that consists of three values, $\ttt$ (true),
$\fff$ (false) and $\uuu$ (unknown).  It is defined by the truth table
\[
\begin{array}{|c|ccc|} \hline
\equiv & \ttt & \fff & \uuu \\ \hline
\ttt & \ttt & \fff & \uuu \\
\fff & \fff & \ttt & \uuu \\
\uuu & \uuu & \uuu & \uuu \\
\hline
\end{array}
\]
The program of \cite{AM01} generates for it 26 minimal
valid membership rules. Out of them 12 are solving rules. For the
remaining rules the sizes of the set $\mathit{friends} \cup
\mathit{obviated}$ are: 17 (for 8 rules), 14 (for 4 rules), and 6 (for
2 rules).

In the \texttt{R} algorithm a selection of a solving rule leads
directly to the termination ($G = \ES$) and to a reduction of the set
$F$ to $\ES$.  For other rules, also a considerable simplification in the
computation takes place. For example, one of the 8 rules with 17 rules in
the set $\mathit{friends} \cup \mathit{obviated}$ is
\[
        r:= x \in \{\fff\}, z \in \{\fff, \uuu\} \;\ra\; y \not= \fff.
\]
Consider now the CSP
$\p{\equiv}{x \in \{\fff\}, y \in \{\fff,\ttt,\uuu\}, z \in \{\fff,\uuu\}}$.
In the \texttt{R} algorithm the selection of $r$ is followed by the
application of the rules from
$\friends$ and the removal of the rules from $\friends \cup \obviated$.
This brings the number of the considered rules down to $26 - 17 = 9$.
The \texttt{R} algorithm subsequently discovers that none of these
nine rules is applicable at this point, so this set $F$ remains upon
termination.  Then in a subsequent constraint propagation phase,
launched after splitting or after constraint propagation involving
another constraint, the fixpoint computation by means of the
\texttt{R} algorithm involves only these nine rules instead of the
initial set of 26 rules. For solving rules this fixpoint computation
immediately terminates.

Interestingly, as Table~\ref{tab:solving} shows, the solving rules
occur quite frequently.  We list there for each constraint and each
type of rules the number of solving rules divided by the total number
of rules, followed in a new line by the average number of rules in the
set $\friends(g) \cup \obviated(g)$.

\begin{table}[htbp]
\begin{tabular*}{\textwidth}{l@{\extracolsep{\fill}}ccccccc}\hline\hline\\[-2ex]
{\bfseries Constraints}
                & \texttt{and2}
                & \texttt{and3}
                & \texttt{and9}
                & \texttt{and11}
                & \texttt{fork}
                & \texttt{rcc8}
                & \texttt{allen}
                \\\hline\\[-2ex]

equality
                & 6/6
                & 13/16
                & 113/134
                & 129/153
                & 9/12
                & 183/183
                & 498/498
                \\
                & 6
                & 14
                & 130
                & 148
                & 11
                & 183
                & 498
                \\[2mm]

membership
                & 6/6
                & 4/13
                & 72/1294
                & 196/4656
                & 0/24
                & 0/912
                & n.a./26446
                \\
                & 6
                & 7
                & 810
                & 3156
                & 9
                & 556
                & n.a.
                \\
\hline\hline
\end{tabular*}
\caption{Solving rules}
    \label{tab:solving}
\end{table}

The \texttt{fork} constraint is taken from the Waltz language for the
analysis of polyhedral scenes. The \texttt{rcc8} is the composition
table for the Region Connection Calculus with 8 relations from
\cite{Ege91}. It is remarkable that all its 183 minimal
valid equality rules are solving.  While none of its 912 minimal valid
membership rule for \texttt{rcc8} is solving, on the average the set
$\friends(g) \cup \obviated(g)$ contains 556 membership
rules.  Also all 498 minimal valid equality rules for the
\texttt{allen} constraint, that represents the composition table for
Allen's qualitative temporal reasoning, are solving. The number of
minimal valid membership rules exceeds 26,000 and consequently they
are too costly to analyze.

\paragraph{Simplification rules.}
\label{para:simplification-rules}
The \texttt{CHR} language supports besides propagation rules also
so-called simplification rules. Using them one can remove constraints
from the constraint store, so one can affect its form.  In \cite{AR01}
a method is discussed that allows one to automatically transform
\texttt{CHR} propagation rules into simplification rules that respects
their semantics.  It is based on identifying or constructing
propagation rules that are solving.

In contrast, our method captures the \emph{degree} to which a rule is solving,
by the ratio of the sizes of $U(r)=\friends(r) \cup \obviated(r)$ and the full rule set.
If the sets are equal, then the ratio is $1$ and $r$ is a solving rule.
Consider now two non-solving rules $r_1,r_2$, that means with
$U(r_1) \subset {\cal R}$ and $U(r_2) \subset {\cal R}$, but
let also \mbox{$U(r_1) \cup U(r_2) = {\cal R}$}.
Suppose that during a fixpoint computation the conditions of both rules have succeeded,
and their bodies have been applied.
The \texttt{R} algorithm would now immediately detect
that the constraint is solved, and consequently terminate.
\texttt{CHR}, for which $r_1$ and $r_2$ are ordinary (propagation) rules,
cannot detect this possibility for immediate termination.

\section{Redundancy of {\prop} Rules}
\label{sec:redundancy}

The cost of a fixpoint computation by the \texttt{GI} algorithm or
one of its derivatives depends on the number of functions involved, in
particular in absence of a good strategy for selecting the functions,
represented in the algorithms by the ``choose'' predicate.  It is
therefore important to identify functions or rules that are not
needed for computing fixpoints.
In the following we shall examine the issue of rule redundancy.
We shall again start with arbitrary functions before moving on to ({\prop}) rules.
The redundancy concept we employ is based on fixpoints.
In the following, for brevity, we drop the word ``common'' when
referring to common fixpoints of a set of functions.

\begin{definition}
\begin{itemize}
\item
Consider a set $F \cup \{f\}$ of functions on a partial ordering.
A function $f$ is called \emph{redundant with respect to $F$}
if the sets of fixpoints of $F$ and $F \cup \{f\}$ are equal.

\item A set of functions $F$ is called \emph{minimal with respect to redundancy}
(or simply \emph{minimal}), if no function $f \in F$ is redundant with respect
to $F - \{f\}$.
\HB
\end{itemize}
\end{definition}
Equivalently, we can say that a function $f$ is redundant w.r.t.\ $F$ if every
fixpoint of $F$ is also a fixpoint of $f$.

\subsection{Redundant Rules}

We now focus on the subject of redundancy for {\prop} rules.
The following simple test is then useful.

\begin{theorem}\label{thm:redrule}
Consider a set  $F$ of {\prop} rules and
a {\prop} rule $r := b \ra g $ with the witness $w$ for $b$.
Let $e$ be the least fixpoint of $F$ greater than or
equal to $w$. If $g(e) = e$, then the rule $r$ is redundant with respect to $F$.
\end{theorem}
\Proof
We show that $g(e) = e$ implies that an arbitrary
fixpoint $d$ of $F$ is a fixpoint of $r$ by a case condition.

\begin{description}
\item [$b$ holds for $d$:]
We have $w \sqsubseteq d$ since $w$ is the witness for $b$.
Also, $w \sqsubseteq e \sqsubseteq d$ since $e$
is the least fixpoint of $F$ greater than or equal to $w$.
From $e \sqsubseteq d$, $g(e) = e$, and the stability of $g$
we conclude $g(d) = d$. Hence $r(d) = (b \ra g)(d) = g(d) = d$.
\item [$b$ does not hold for $d$:]
Then $r(d) = (b \ra g)(d) = d$.
\HB
\end{description}

This test is of interest to us since it allows us to compute only one fixpoint
of $F$ instead of all fixpoints.  It is effective if
\begin{itemize}
\item  the witness can be computed,
\item  the equality $g(e) = e$ can be determined, and
\item  the fixpoint computations are effective.
\end{itemize}

For the sake of fixpoint computations a rule $r = b \ra g$ with a body
$g = g_1, \LL, g_n$ (describing a function composition) such that any two
different $g_i, g_j$ commute can be identified with the collection $b
\ra g_1, \ldots, b \ra g_n$ of the rules, and vice versa.  Indeed, the
respective fixpoints and the rule properties are the same.  We
consider here these two representations as equivalent.  If a rule with
such a ``compound'' body is not redundant it might be so in part.
That is, some part of its body might be redundant or, in other words,
some sub-rules of its decomposition might be.  This is what we mean
below by saying that a rule is \emph{partially} redundant.

Let us consider now the task of computing minimal sets of {\prop}
rules.  Such sets can of course be generated by a simple bounded loop:
select an untested rule, test whether it is redundant and, if so,
remove it from the current set.  In general, however, the obtained
minimal sets depend on the selection order for testing; see an example
below.  In our experiments we used a strategy that selects first the
rules the execution of which is computationally expensive, for
instance due to conditions on many variables.  In this way we hope to
obtain a set of computationally cheap rules.

\subsection{An Example: Redundant Membership Rules}
\label{sec:example}

Let us illustrate now a number of issues by means of an example.
Consider the constraint $c(x,y,z,u)$
defined by
\[
\begin{array}{|cccc|} \hline
x & y & z & u \\ \hline
0 & 1 & 0 & 1 \\
1 & 0 & 0 & 1 \\
1 & 1 & 1 & 0 \\
\hline
\end{array}
\]
The underlying domain for all its variables is $\{0,1\}$.
Hence the induced corresponding partial order is
\[
        (\C{(A,B,C,D) \mid A,B,C,D \sse \C{0,1}}, \:\supseteq).
\]
The algorithm of \cite{AM01} generates eleven membership rules listed
in Figure~\ref{fig:cr}.
Since the rule conditions are only equality tests,
we use an alternative notation that should  be self-explanatory.

\newcounter{xsetcountertmpstore}
\newcommand{\xsetcountertmp}[2] {\setcounter{xsetcountertmpstore}{\value{#1}}\setcounter{#1}{#2}}
\newcommand{\xresetcounter}[1]  {\setcounter{#1}{\value{xsetcountertmpstore}}}

\begin{figure}[ht]
\xsetcountertmp{equation}{0}
\begin{eqnarray}
c(x, y, z, 0)  & \quad\ra\quad & x \neq 0, y \neq 0, z \neq 0\\
c(x, y, 1, u)  & \ra &  u \neq 1, \underline{x \neq 0, y \neq 0}\\
c(0, y, z, u)  & \ra &  u \neq 0, y \neq 0, \underline{z \neq 1}\\
c(x, 0, z, u)  & \ra &  u \neq 0, x \neq 0, \underline{z \neq 1}\\
c(x, y, z, 1)  & \ra &  z \neq 1\\
c(x, y, 0, u)  & \ra &  u \neq 0\\
c(1, 1, z, u)  & \ra &  u \neq 1, \underline{z \neq 0}\\
c(x, 1, 0, u)  & \ra &  x \neq 1\\
c(x, 1, z, 1)  & \ra &  \underline{x \neq 1}\\
c(1, y, 0, u)  & \ra &  y \neq 1\\
c(1, y, z, 1)  & \ra &  \underline{y \neq 1}
\end{eqnarray}
\xresetcounter{equation}
\caption{Membership rules for the constraint $c$}
\label{fig:cr}
\end{figure}

For example, rule (11) states that if $c(x, y, z, u)$,
then it is correct to conclude from $x=1$ and $u=1$
that $y \neq 1$ (validity), and furthermore that neither $x=1$ nor
$u=1$ suffices for this conclusion (minimality).

Suppose we are interested
in computing the smallest fixpoint greater than or equal to
$
        E_1 = \{1\} \times \{0,1\} \times \{0,1\} \times \{1\}.
$
Suppose rule (11) is considered.  Its application yields
$
        E_2 = \{1\} \times \{0\} \times \{0,1\} \times \{1\}
$
from where rule (4) leads to
\mbox{$
        E_3 = \{1\} \times \{0\} \times \{0\} \times \{1\} .
$}
This is indeed a fixpoint since for each rule either its condition does not
apply or the application of its body results again in $E_3$.

A second possible iteration from $E_1$ that stabilises in $E_3$ is
by rule (5) followed by rule (10).  Rule (11) can be applied at this point
but its body does not change $E_3$. Indeed,
$E_3$ is a fixpoint of all rules including rule (11).
We conclude that rule~(11) is redundant --
we just performed the test of Theorem~\ref{thm:redrule}.

The process of identifying redundant rules can then be continued for
the rule set $\{(1), \ldots, (10)\}$.  One possible outcome is depicted in
Figure~\ref{fig:cr}, where redundant parts of rules are underlined.
From the 20 initial atomic conclusions 13 remain, thus
we find here a redundancy ratio of 35\%.

Consider now the justification for the redundancy of rule (11), and observe
that rule (11) has no effect since rule (10), which has the same body, was
applied before.  Suppose now that the process of redundancy
identification is started with rule (10) instead of rule (11).  This
results in identifying rule (10) as redundant, with a relevant
application of rule (11).

Note moreover that one of the rules (10), (11) must be present
in \emph{any} minimal set since their common body
$y \neq 1$ occurs in no other rule.  It would seem difficult to find
a criterion that prefers one rule over the other as their structure is the same.

\subsection{Experiments}

We implemented in {\eclipse} an algorithm that computes
minimal sets of membership rules.
The results for some benchmark rule sets are listed
in Table~\ref{table:reduexp}.

\begin{table}[ht]
\begin{tabular*}{\textwidth}{l@{\extracolsep{\fill}}ccccccc}
\hline\hline\\[-3mm]
                & \texttt{and11}$_M$
                & \texttt{and11}$_E$
                & \texttt{and3}$_M$
                & \texttt{equ3}$_M$
                & \texttt{fula2}$_E$
                & \texttt{fork}$_E$
                & \texttt{fork}$_M$
                \\\hline\\[-2ex]
total
                & 4656
                & 153
                & 18
                & 26
                & 52
                & 12
                & 24
                \\[1mm]
\parbox{15mm}{redundant\\[-0.5mm](partially)}
                & 4263 (2)
                & 0 (6)
                & 5 (0)
                & 8 (0)
                & 24 (0)
                & 0 (9)
                & 6 (6)
                \\[3mm]
\parbox{13mm}{redundancy\\[-0.5mm]ratio}
                & 81\%
                &  4\%
                & 30\%
                & 26\%
                & 35\%
                & 35\%
                & 40\%
                \\[1mm]
\hline\hline\\[-1ex]
\end{tabular*}
\caption{Minimizing rule sets}
\label{table:reduexp}
\end{table}
For each constraint the set of minimal membership or equality rules
(indicated respectively by the subscript ``$_M$'' or ``$_E$'')
was computed by the rule generation algorithm of
\cite{AM01}.
The constraints are taken from the experiments discussed in Table
\ref{tab:fixpoints}. Additionally a 5-ary constraint \texttt{fula}
(standing for the well-known \texttt{fulladder} constraint)
is analyzed.

The table shows the size of the rule set, the number of fully and, in
parentheses, partially redundant rules.
The redundancy ratio for the entire rule set shows the
percentage of the atomic disequalities that are removed
from the rule conclusions on the account of redundancy.

Computation times are negligible in so far as they are considerably
smaller than the corresponding rule generation times.

\subsection{Schedulers and Minimal Rule Sets}
\label{sec:schedulers-redundancy}

There is no simple connection
between redundancy and the rule sets $\friends$ and $\obviated$
of the \texttt{R} scheduler.
For instance, it is \emph{not} the case that
a rule is redundant if it is contained in $\friends(r)\cup\obviated(r)$ of every rule $r$.
Nor is a redundant rule necessarily contained in $\friends(r)\cup\obviated(r)$ of every rule $r$.
To examine this in an example, recall the rules in Figure~\ref{fig:cr}.
All except (5) and (6) are solving rules, i.e., each
respective set $\friends\cup\obviated$ is the complete set $\{(1),\ldots,(11)\}$
of rules, while for rules (5) and (6) this set is $\{(1),(3),(5),(6)\}$.
Further, neither (5) nor (6) is redundant with respect to all other rules,
whereas (10) and (11) are.

\paragraph{Benchmarks.}
We reran the benchmarks from Tables~\ref{tab:fixpoints} and \ref{tab:atpg}
with all involved rule sets subjected to a removal of redundant rules
and subsequent recomputation of the sets $\friends$ and $\obviated$.
The results are reported in Tables~\ref{tab:fixpoints:nr} and \ref{tab:atpg:nr} below.
The rule sets of \texttt{rcc8} were already minimal; therefore this constraint is omitted.

\begin{table}[ht]
  \begin{center}
\begin{tabular*}{\textwidth}{l@{\extracolsep{\fill}}ccccc}\hline\hline\\[-2ex]
{\bfseries Constraint}
        & \texttt{fork}
        & \texttt{and3}
        & \texttt{and9}
        & \texttt{and11}
        \\\hline
\textsc{membership}\\[1ex]
\quad relative
        & 60\% / 46\%
        & 69\% / 48\%
        & 28\% / 18\%
        & 50\% / 29\%
        \\[1mm]
\quad absolute
        & {\small 0.32/0.53/0.70}
        & {\small 0.27/0.39/0.56}
        & {\small 167/589/924}
        & {\small 157/316/543}
        \\[2mm]
\textsc{equality}\\[1ex]
\quad relative
        & 97\% / 93\%
        & 97\% / 64\%
        & 96\% / 101\%
        & 96\% / 101\%
        \\[1mm]
\quad absolute
        & {\small 21.6/22.2/23.2}
        & {\small 0.37/0.38/0.58}
        & {\small 386/404/384}
        & {\small 341/353/339}
\\\hline\hline\\[-1ex]
\end{tabular*}
\caption{Randomized search trees for single constraints (without redundant rules)}
\label{tab:fixpoints:nr}
\end{center}
\end{table}

\begin{table}[ht]
\begin{center}
\begin{tabular*}{\textwidth}{l@{\extracolsep{\fill}}cccc}\hline\hline\\[-2ex]
{\bfseries Logic}
        & 3-valued
        & 9-valued
        & 11-valued
        \\\hline
\textsc{membership}\\[1ex]
\quad relative
        & 66\% / 46\%
        & 62\% / 33\%
        & 68\% / 35\%
        \\[1mm]
\quad absolute
        & {\small 1.32/2.00/3.05}
        & {\small 37/59/114}
        & {\small 70/103/199}
        \\[2mm]
\textsc{equality}\\[1ex]
\quad relative
        & 61\% / 26\%
        & 40\% / 58\%
        & 33\% / 48\%
        \\[1mm]
\quad absolute
        & {\small 0.72/1.18/2.73}
        & {\small 2.57/6.41/4.46}
        & {\small 13.8/41.0/28.6}
\\\hline\hline\\[-1ex]
\end{tabular*}
\caption{CSPs formalizing sequential ATPG (without redundant rules)}
\label{tab:atpg:nr}
\end{center}
\end{table}

When comparing the redundancy and non-redundancy benchmarks versions
we observe that the absolute execution times are enormously reduced
in the case of the constraints on higher-valued logics.
This is in line with the much smaller sizes of the reduced rule sets.
The ratios of the execution times, however, are barely affected.
The type of a scheduler and minimality w.r.t.\ redundancy
appear to be rather orthogonal issues.

\smallskip

It is interesting to examine in one case the distribution of the solving degrees,
i.e., the ratios of the sizes of $\friends \cup \obviated$
and the full rule set.
Recall that a ratio of $1$ means that the
constraint is solved when the rule body has been executed.
Such a rule could be represented as a simplification rule in \texttt{CHR}
(see Section~\ref{para:simplification-rules}).

\begin{figure}[ht]
\includegraphics[width=0.36\columnwidth,angle=-90]{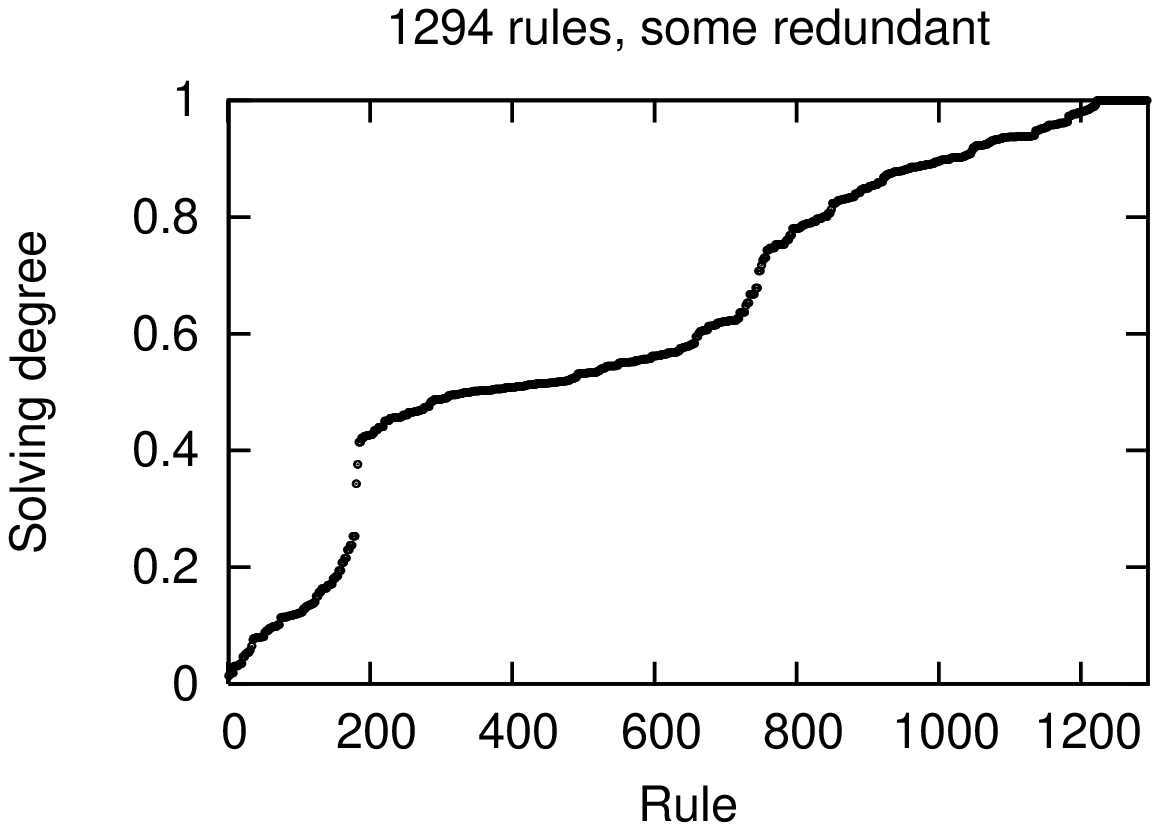}%
\includegraphics[width=0.36\columnwidth,angle=-90]{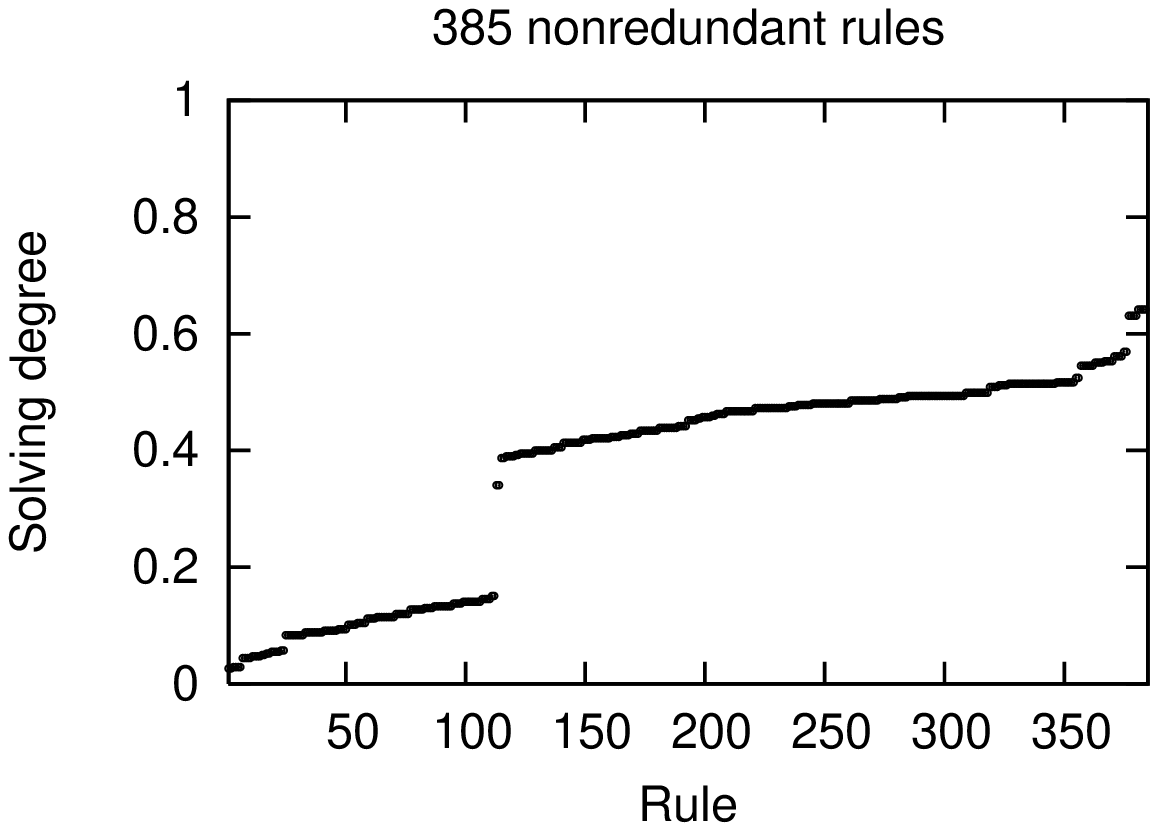}
\caption{\texttt{and9$_M$}: Solving degree and redundancy}
\label{fig:solving-degree}
\end{figure}

In Figure~\ref{fig:solving-degree} two membership rule sets
for the constraint \texttt{and9} are compared. One set contains
redundant rules, the other set is minimal w.r.t.\ redundancy.
The rules in the minimal set are solving to a lesser degree.
In particular, none is a proper solving rule.
The good performance of the \texttt{R} algorithm in the benchmarks of
Tables~\ref{tab:fixpoints:nr},\ref{tab:atpg:nr} may thus be attributed not
to distinguishing solving (simplification) rules and
non-solving propagation rules, but to the \emph{accumulated} effect
of removing rules from the fixpoint computation.

\section*{Acknowledgments}
We thank Christian Holzbaur and Eric Monfroy for helpful discussions
on the implementation and on an early version of this paper.
The reviewers made useful comments which helped to improve the paper.


\begin{thebibliography}{}

\bibitem[\protect\citeauthoryear{Abdennadher, {Kr\"{a}mer}, Saft, and
  Schmaus}{Abdennadher et~al\mbox{.}}{2001}]{AKSS01}
{\sc Abdennadher, S.}, {\sc {Kr\"{a}mer}, E.}, {\sc Saft, M.}, {\sc and} {\sc
  Schmaus, M.} 2001.
\newblock {JACK: A Java Constraint Kit}.
\newblock In {\em International Workshop on Functional and (Constraint) Logic
  Programming ({WFLP 2001}), Technical Report No. 2017}. University of Kiel,
  Kiel, Germany.

\bibitem[\protect\citeauthoryear{Abdennadher and Rigotti}{Abdennadher and
  Rigotti}{2001}]{AR01}
{\sc Abdennadher, S.} {\sc and} {\sc Rigotti, C.} 2001.
\newblock Using confluence to generate rule-based constraint solvers.
\newblock In {\em Proceedings of the 3rd Int. Conf. on Principles and Practice
  of Declarative Programming ({PPDP 2001})}. ACM, Firenze, Italy, 127--135.

\bibitem[\protect\citeauthoryear{Apt}{Apt}{1999}]{Apt99b}
{\sc Apt, K.~R.} 1999.
\newblock The essence of constraint propagation.
\newblock {\em Theoretical Computer Science\/}~{\em 221,\/}~1--2, 179--210.
\newblock Available via \verb+http://arXiv.org/archive/cs/+.

\bibitem[\protect\citeauthoryear{Apt}{Apt}{2000}]{Apt00a}
{\sc Apt, K.~R.} 2000.
\newblock The role of commutativity in constraint propagation algorithms.
\newblock {\em ACM Transactions on Programming Languages and Systems\/}~{\em
  22,\/}~6, 1002--1036.
\newblock Available via \verb+http://arXiv.org/archive/cs/+.

\bibitem[\protect\citeauthoryear{Apt and Monfroy}{Apt and Monfroy}{2001}]{AM01}
{\sc Apt, K.~R.} {\sc and} {\sc Monfroy, E.} 2001.
\newblock Constraint programming viewed as rule-based programming.
\newblock {\em Theory and Practice of Logic Programming\/}~{\em 1,\/}~6,
  713--750.
\newblock Available via \verb+http://arXiv.org/archive/cs/+.

\bibitem[\protect\citeauthoryear{Brand}{Brand}{2001}]{brand:atpg:2001}
{\sc Brand, S.} 2001.
\newblock Sequential automatic test pattern generation by constraint
  programming.
\newblock In {\em CP2001 Post Conference Workshop Modelling and Problem
  Formulation}.
\newblock Available via \verb+http://homepages.cwi.nl/~sbrand/+.

\bibitem[\protect\citeauthoryear{Egenhofer}{Egenhofer}{1991}]{Ege91}
{\sc Egenhofer, M.} 1991.
\newblock Reasoning about binary topological relations.
\newblock In {\em Proceedings of the 2nd International Symposium on Large
  Spatial Databases ({SSD})}, {O.~G{\"{u}}nther} {and} {H.-J. Schek}, Eds.
  Lecture Notes in Computer Science, vol. 525. Springer-Verlag, 143--160.

\bibitem[\protect\citeauthoryear{Fr{\"{u}}hwirth}{Fr{\"{u}}hwirth}{1998}]{Frue%
hwirthJLP98}
{\sc Fr{\"{u}}hwirth, T.} 1998.
\newblock Theory and practice of {Constraint Handling Rules}.
\newblock {\em Journal of Logic Programming\/}~{\em 37,\/}~1--3 (October),
  95--138.
\newblock Special Issue on Constraint Logic Programming (P.~J. Stuckey and K.
  Marriot, Eds.).

\bibitem[\protect\citeauthoryear{Holzbaur, de~la Banda, Jeffery, and
  Stuckey}{Holzbaur et~al\mbox{.}}{2001}]{Holzbaur:2001:OCC}
{\sc Holzbaur, C.}, {\sc de~la Banda, M.~G.}, {\sc Jeffery, D.}, {\sc and} {\sc
  Stuckey, P.~J.} 2001.
\newblock Optimizing compilation of constraint handling rules.
\newblock In {\em Proceedings of the 2001 International Conference on Logic
  Programming}. Lecture Notes in Computer Science, vol. 2237. Springer-Verlag,
  74--89.

\bibitem[\protect\citeauthoryear{Kleene}{Kleene}{1952}]{Kle52}
{\sc Kleene, S.~C.} 1952.
\newblock {\em Introduction to Metamathematics}.
\newblock van Nostrand, New York.

\end{thebibliography}

\end{document}